\documentclass[10pt, conference]{IEEEtran}
\IEEEoverridecommandlockouts

\usepackage{cite}
\usepackage{amsmath,amssymb,amsfonts}
\usepackage{algorithmic}
\usepackage{graphicx}
\usepackage{textcomp}
\usepackage{tcolorbox}
\usepackage[ruled,vlined,linesnumbered]{algorithm2e}
\usepackage{graphicx}
\usepackage{hyperref}
\usepackage{float}
\usepackage{subfigure}
\usepackage[subfigure]{tocloft}
\usepackage{booktabs}
\usepackage{multirow}
\usepackage{color}
\usepackage{xcolor}
\usepackage{tabularx}
\usepackage{url}
\usepackage{amsthm}
\newtheorem{example}{Example}
\newtheorem{definition}{Definition}

\def\BibTeX{{\rm B\kern-.05em{\sc i\kern-.025em b}\kern-.08em
    T\kern-.1667em\lower.7ex\hbox{E}\kern-.125emX}}
\begin{document}

\newtcbox{\mybox}[1][red]
  {on line, arc = 0pt, outer arc = 0pt,
    colback = #1!10!white, colframe = #1!50!black,
    boxsep = 0pt, left = 1pt, right = 1pt, top = 2pt, bottom = 2pt,
    boxrule = 0pt, bottomrule = 1pt, toprule = 1pt}

\title{Improving NLSAT for Nonlinear Real Arithmetic}

\author{
    \IEEEauthorblockN{Zhonghan Wang\IEEEauthorrefmark{1}\IEEEauthorrefmark{2}}
    \IEEEauthorblockA{\IEEEauthorrefmark{1}Key Laboratory of System Software (Chinese Academy of Sciences) and State Key \\ Laboratory of
 Computer Science, Institute of Software, Chinese Academy of Sciences, Beijing, China}
    \IEEEauthorblockA{\IEEEauthorrefmark{2}University of Chinese Academy of Sciences, Beijing, China\\
    wangzh@ios.ac.cn, wangzhonghan272@gmail.com}
    \thanks{This work was conducted during the author's Master studies at the Institute of Software, Chinese Academy of Sciences.}
}

\maketitle

\begin{abstract}
The Model-Constructing Satisfiability Calculus (MCSAT) framework has been applied to SMT problems over various arithmetic theories. NLSAT, an implementation using cylindrical algebraic decomposition (CAD) for explanation, is especially competitive for nonlinear real arithmetic (NRA) constraints. However, current Conflict-Driven Clause Learning (CDCL)-style algorithms only consider literal information when making decisions, and thus ignore the influence of clauses on arithmetic variables. This limitation may lead NLSAT to encounter unnecessary conflicts due to suboptimal literal choices. To address this issue, we analyze conflicts caused by literal decisions and incorporate clause-level information that directly affects arithmetic variables. We propose two main algorithmic improvements: a clause-level feasible-set-based look-ahead mechanism and an arithmetic propagation-based branching heuristic. We implement our solver, named clauseSMT, based on a dynamic variable ordering framework. Experiments indicate that clauseSMT is competitive on nonlinear real arithmetic problems compared with existing SMT solvers (CVC5, Z3, YICES2), and it outperforms all of them on satisfiable instances of SMT(QF\_NRA) in SMT-LIB. We also evaluate the effectiveness of our proposed methods.

\end{abstract}

\begin{IEEEkeywords}
NLSAT, nonlinear real arithmetic, SMT, clause level.
\end{IEEEkeywords}

\section{Introduction}
\subsection{Motivation}
Satisfiability Modulo Theories (SMT) refers to the problem of determining the satisfiability of formulas in first-order logic. SMT problems typically involve theories such as linear and nonlinear arithmetic, uninterpreted functions, strings, and arrays~\cite{BarrettT18}. As a fundamental problem in software engineering, formal methods, and programming languages, SMT has widespread applications, including symbolic execution~\cite{KLEE, DART}, program verification~\cite{AnalysisSymbol, VerificationSMT}, program synthesis~\cite{synthesis1}, automata learning~\cite{Automata1, automata2}, and neural network verification~\cite{NN1, NN2, NN3, NN4}.

Nonlinear real arithmetic (NRA) is a class of arithmetic theories. It consists of atoms represented as inequalities over polynomials, and is therefore sometimes referred to as the theory of polynomial constraints. Variables can take Boolean or real values, depending on their types. SMT(NRA) instances are typically generated from both academic and industrial applications. They are commonly used in cyber-physical systems~\cite{CPS1, CPS2, CPS3}, ranking function generation~\cite{LeikeH15, HeizmannHLP13}, and nonlinear hybrid automata analysis~\cite{CimattiMT12}. Instances from these applications are collected in the SMT-LIB benchmarks~\cite{BarFT-SMTLIB}. The high performance of SMT solvers over nonlinear arithmetic has significantly improved these applications.

Decision procedures for solving nonlinear arithmetic are usually based on cylindrical algebraic decomposition (CAD)\cite{Caviness2004QuantifierEA}, a widely used tool for real quantifier elimination. CAD generates the current unsatisfiable cell during the search procedure and is employed in modern algorithms. Among these, NLSAT\cite{JovanovicM12} is a mainstream algorithm that leverages CAD for lemma generation. Its core idea is to assign values directly to arithmetic variables, rather than at the literal level as in CDCL(T).

Although NLSAT introduces the novel approach of directly assigning arithmetic variables, it still relies on literal decisions when processing arithmetic clauses containing several unevaluated literals. These decided literals are then used for conflict analysis within the CDCL-style framework. However, improper literal decisions can sometimes induce conflicts, slowing down the overall search process. Therefore, a heuristic for literal decisions is necessary.

We identify three central problems and present our solutions in the context of algorithmic improvements:
\begin{itemize}
    \item What factors cause conflicts in the NLSAT algorithm, and can some of them be avoided?
    \item Is it possible to assign values directly to arithmetic variables, independent of literal-level decision information, within a CDCL-style framework?
    \item Can propagation be performed on arithmetic variables, analogous to unit propagation in SAT solving, and is this new propagation method effective for guiding assignments and detecting conflicts?
\end{itemize}

\subsection{Contributions}
To address the questions above, this paper proposes, for the first time, a new algorithm that incorporates clause-level information.

First, we analyze the conflict problems that arise in NLSAT and categorize them into two types. As described in~\cite{multilinear}, each clause narrows the feasible set\footnote{Also called the satisfying domain in~\cite{multilinear}.} of an arithmetic variable. This technique has previously been used to enlarge the operation choices in local search algorithms, but it has not been considered in complete methods like NLSAT. Consequently, arithmetic variables can sometimes be narrowed to an empty search space, causing conflicts. We describe this type of problem from the perspective of interval arithmetic and propose a solution based on the computation of feasible intervals. The clause-level feasible-set idea extends the spirit of NLSAT by directly guiding assignments to arithmetic variables.

Second, we introduce an incremental computation of the clause-level feasible set, followed by the definition of \textbf{clause-level propagation}. In SAT solvers, unit propagation is an effective tool to deduce assignments and detect conflict clauses. Analogously, clause-level propagation is employed to fix a possible witness for an arithmetic variable or to quickly detect empty feasible-set cases.

Finally, we present the structure and implementation details of our solver, \texttt{clauseSMT}. Although dynamic variable ordering has been discussed in~\cite{MCSATOrder}, SMT-RAT~\cite{smtrat} solves fewer instances due to the lack of efficient data structures. We present techniques and data structures inspired by SAT solving. Our implementation extends the NLSAT module of the Z3 solver~\cite{MouraB08}, relying on existing libraries in Z3 for mathematical operations such as root isolation, polynomial operations, and algebraic number representation. Experiments on the SMT-LIB benchmark demonstrate the effectiveness of our proposed techniques, including the look-ahead mechanism and clause-level propagation. The results show that \texttt{clauseSMT} solves the most satisfiable instances and is highly competitive against other SMT solvers overall.

In summary, this paper makes the following contributions:
\begin{itemize}
    \item We propose a new MCSAT-based method for nonlinear arithmetic, introducing \textbf{clause-level feasible sets} to avoid conflicts caused by literal-level decisions.
    \item We define \textbf{clause-level propagation}, which quickly detects conflict cases or fixes values for arithmetic variables.
    \item We integrate the propagation method into the VSIDS branching heuristic, guiding the search process and reducing semantic stages.
    \item We implement these ideas in our solver \texttt{clauseSMT} and conduct experiments on SMT-LIB benchmarks to demonstrate the effectiveness of our approach.
\end{itemize}

\subsection{Structure of the Paper}
The paper is organized as follows. In Section~\ref{sec:pre}, we introduce SMT problems over nonlinear real arithmetic and review the traditional complete method NLSAT. Section~\ref{sec:conflict} analyzes the conflicts that occur in the NLSAT algorithm. In Section~\ref{sec:look-ahead}, we present a feasible-set based look-ahead mechanism. Building on the concept of clause-level information, Section~\ref{sec:prop} introduces the clause-level propagation algorithm and a new branching heuristic. Section~\ref{sec:imp} discusses the details of implementation. We compare our solver with other SMT solvers and perform an ablation study in Section~\ref{sec:eva}. Related work on solving non-linear real arithmetic is reviewed in Section~\ref{sec:related}. Finally, Section~\ref{sec:conclu} concludes the paper and outlines potential directions for future research.

\subsection{Artifact Availability}
To facilitate reproducibility and further research, we release the full implementation of ClauseSMT, together with all experimental data and scripts, as an open-source artifact. The artifact is publicly available on a GitHub repository\footnote{\url{https://github.com/yogurt-shadow/ClauseSMT_ASE2025}}, enabling researchers to replicate our experiments and investigate the solver's performance on the QF\_NRA benchmark.

\section{Preliminaries}
\label{sec:pre}
This section introduces the basic definitions of SMT problems over nonlinear real arithmetic, followed by a review of the NLSAT algorithm. In addition, we present the computation of clause-level feasible sets.

\subsection{Syntax of SMT(QF\_NRA)}
The syntax of SMT constraints over nonlinear real arithmetic is defined as follows:
\begin{align*}
   \mbox{arithmetic variables:} &\quad x \in \mathbb{V}\\
   \mbox{boolean variables:} &\quad b \in \mathbb{B} \\
   \mbox{polynomials:} &\quad p := x ~|~ c ~|~ p + p ~|~ p \cdot p \\
   \mbox{atoms:} &\quad a := b ~|~ p \le 0 ~|~ p \ge 0 ~|~ p = 0 \\
   \mbox{literals:} &\quad l := a ~|~ \neg a \\
   \mbox{formulas:} &\quad \varphi := l ~|~ \varphi \vee \varphi ~|~ \varphi \wedge \varphi
\end{align*}

An atom is either a Boolean atom, defined by a Boolean variable $b \in \mathbb{B}$, or an arithmetic atom, defined by a (non-strict) inequality or equality of a polynomial over $\mathbb{V}$. 
A literal is either an atom or its negation. A clause is a disjunction of literals, and all input formulas are transformed into conjunctive normal form (CNF), i.e., a conjunction of clauses.
SMT(NRA) refers to the set of formulas over the theory of nonlinear real arithmetic.

For the semantics, we define an \emph{assignment} $\alpha$ as a mapping from variables to values. 
\begin{itemize}
  \item A \emph{Boolean assignment} maps Boolean variables to truth values, denoted as 
  $\alpha_{\mathsf{bool}}: b \mapsto \{\top,\bot\}$.  
  \item An \emph{arithmetic assignment} maps real variables to real numbers, denoted as 
  $\alpha_{\mathsf{real}}: x \mapsto \mathbb{R}$.  
\end{itemize}

A \emph{full assignment} maps all Boolean and real variables, while a \emph{partial assignment} only covers a subset. 
Under a given assignment, each atom is evaluated as follows: 
\begin{enumerate}
  \item \emph{true}, if the assignment satisfies it;  
  \item \emph{false}, if the assignment violates it;  
  \item \emph{undefined or unevaluated}, if it contains variables not yet assigned.  
\end{enumerate}

A full assignment that makes all clauses true is called a \emph{model} (or \emph{solution}) of the formula, certifying its satisfiability. 
The SMT(QF\_NRA) problem is to decide whether such a model exists for a given input formula, or to prove that none does.

\subsection{Feasible Set}
For nonlinear real arithmetic constraints, a key technique for determining the possible values of arithmetic variables is \emph{root isolation}.  
Given an arithmetic atom of the form 
\[
  p \ \{\le, \ge, =, >, <\} \ 0,
\] 
if exactly one variable $v$ remains unassigned under the current assignment, we can compute the set of values of $v$ that satisfy the atom. We call this set the \emph{feasible set}.  

For higher-order polynomial constraints, feasible sets are usually computed via root isolation, which determines the roots of the polynomial. These roots partition the real line into intervals\footnote{In the terminology of cylindrical algebraic decomposition, such intervals are also called \emph{cells}.}.  
Each interval preserves a fixed truth value of the atom, and the union of all satisfying intervals forms the overall feasible set.  

For negated literals, the feasible set is simply the complement of the feasible set of the corresponding positive atom. By restricting the variable $v$ to any value in the feasible set, the atom is guaranteed to be satisfied\footnote{The feasible set may also be empty or cover the entire real line, meaning the atom is always unsatisfiable or always satisfied, respectively.}.

Besides the literal level, the notion of a feasible set can also be extended to the clause level, where it represents the set of values that make the entire clause satisfied. In this work, we focus on the case where exactly one arithmetic variable in the clause is left unassigned; we call such a clause a \emph{univariate clause}.  

\begin{definition}[Feasible Set]
Let $l$ be a literal, $x$ an arithmetic variable, and $\alpha$ an assignment that maps all variables in $l$ except $x$.  
The \emph{feasible set} (resp. \emph{infeasible set}) of $l$ is the union of intervals over $\mathbb{R}$ such that $l$ is satisfied (resp. unsatisfied) when $x$ is assigned any value from the interval.  

Similarly, let $c$ be a clause, $x$ an arithmetic variable, and $\alpha$ an assignment that maps all variables in $c$ except $x$.  
The \emph{feasible set} (resp. \emph{infeasible set}) of $c$ is the set of all values of $x$ that make $c$ satisfied (resp. unsatisfied).  
It can be computed by taking the union (resp. intersection) of the feasible sets (resp. infeasible sets) of all literals in $c$.  

Example~\ref{ex:feasible_set} illustrates the construction of a feasible set.
\end{definition}

\begin{example}
\label{ex:feasible_set}
Consider an assignment $\alpha := \{ b \mapsto \bot, x \mapsto 0 \}$.  
The feasible set of the clause
\[
b \vee (y + x > 0) \vee (y^2 > 2)
\]
is 
\[
(-\infty, -\sqrt{2}) \ \cup\ (0, \infty),
\]
since these are the values of $y$ that satisfy the clause given the current assignment.
\end{example}

\subsection{Original NLSAT Algorithm}
NLSAT is the core algorithm for nonlinear real arithmetic (NRA) within the Z3 solver~\cite{MouraB08}. It handles both boolean and arithmetic variables directly, integrating theory reasoning into the CDCL framework. Specifically, NLSAT extends traditional unit propagation and boolean decisions to real-variable propagation (R-propagation) and semantic decisions.

To select an appropriate value for an arithmetic variable, NLSAT incrementally updates its feasible-set during the search. Let the current feasible-set be $curr\_set$, the feasible-set of a literal $lit$ be $lit\_set$, and real-variable propagation take effect under the following circumstances:
\begin{itemize}
    \item $\mathbf{lit\_set}$ \textbf{is empty}: the literal is propagated as false, since no value can satisfy it.
    \item $\mathbf{lit\_set}$ \textbf{is full}: the literal is propagated as true, since any value satisfies it.
    \item $\mathbf{curr\_set}$ \textbf{is a subset of} $\mathbf{lit\_set}$: the literal is propagated as true, because the current feasible-set already satisfies it.
    \item $\mathbf{curr\_set}$ \textbf{has no intersection with} $\mathbf{lit\_set}$: the literal is propagated as false, because all values in the current feasible-set violate the literal.
\end{itemize}
When processing a clause containing both boolean and arithmetic variables, NLSAT first applies propagation and evaluation to detect evaluated literals. If one literal is true, the clause is skipped; otherwise, NLSAT decides the first unevaluated literal and updates the feasible-set accordingly. 
Algorithm~\ref{alg:nlsat_process} presents the detailed processing steps in NLSAT.

\begin{algorithm}[!t]
\caption{Clause Processing in NLSAT}
\label{alg:nlsat_process}
\SetKwInOut{Input}{Input}
\SetKwInOut{Output}{Output}

\Input{A set of clauses $F$}
\Output{Conflict clause $conf\_cls$, or \texttt{No Conflict}}

\For{each clause $c \in F$}{
    \For{each literal $l \in c$}{
        $lit\_set \gets$ compute feasible-set of $l$\;
        $val \gets$ real propagate $l$ using $lit\_set$\;
        \If{$val = \top$}{
            break\;  \tcp{clause is satisfied, skip remaining literals}
        }
        \If{$val = \bot$}{
            continue\;  \tcp{literal unsatisfied, check next literal}
        }
    }
    \If{exist satisfied literal in $c$}{
        continue\;  \tcp{clause satisfied, check next clause}
    }
    \ElseIf{exactly one literal undefined in $c$}{
        unit propagate the literal\;
    }
    \ElseIf{two or more literals undefined in $c$}{
        decide the first undefined literal\;
    }
    \Else{ \tcp{all literals are unsatisfied}
        \Return{$c$}  \tcp{conflict detected}
    }
}
\Return{No Conflict}\;
\end{algorithm}

For conflict analysis, NLSAT employs cylindrical algebraic decomposition (CAD) as an explanation tool. Using model-based projection, it identifies the conflict cell and generates a lemma to prevent the solver from revisiting the same conflict in the future. 
Algorithm~\ref{alg:nlsat} presents the complete NLSAT procedure.

\begin{algorithm}[!t]
\caption{Original NLSAT}
\label{alg:nlsat}
\SetKwInOut{Input}{Input}
\SetKwInOut{Output}{Output}

\Input{A formula $F$}
\Output{SAT or UNSAT}

\While{true}{
    $v \gets$ select next variable according to branching heuristic\;
    $conf\_cls \gets$ process clauses univariate to $v$ (Algorithm~\ref{alg:nlsat_process})\;
    
    \If{$conf\_cls$ is empty}{
        \tcp{No conflict detected}
        \If{$v$ is boolean}{
            perform boolean decision\;
        }
        \ElseIf{$v$ is arithmetic}{
            perform semantic decision\;
        }
        \Else{
            \Return{SAT} \tcp{all variables assigned consistently}
        }
    }
    \Else{
        \tcp{Conflict detected}
        $new\_lemma \gets$ conflict analysis via CAD\;
        \If{$new\_lemma$ is empty}{
            \Return{UNSAT} \tcp{formula is unsatisfiable}
        }
        \Else{
            backtrack\;
        }
    }
}
\end{algorithm}

\section{Conflicts During the NLSAT Algorithm}
\label{sec:conflict}

In this section, we analyze the sources of conflicts in NLSAT algorithms. Broadly, these conflicts can be categorized into two types: those caused by semantic decisions and those caused by literal decisions.

\subsection{Conflicts Caused by Semantic Decisions}

In SAT solving, conflicts are typically caused by literal (i.e., boolean variable) decisions. For a satisfiable instance, a SAT solver can avoid conflicts if it uses a perfect phase selection strategy (i.e., assigns variables correctly to true or false). For unsatisfiable instances, conflicts are unavoidable regardless of the phase choices. In both cases, when CDCL detects a conflict due to incorrect decision values, conflict analysis is invoked to generate a new lemma that forces a change in the previous assignment.

Similarly, in the NLSAT algorithm for SMT solving, conflicts can arise from incorrect semantic decisions, i.e., selecting a value from a given interval. As discussed in~\cite{LiXZ23}, the search space of nonlinear arithmetic is partitioned into sign-invariant cells. However, in systematic solvers like NLSAT, the current cell being explored cannot be predicted in advance, and therefore conflicts may occur. For unsatisfiable instances, every cell in the search space is inconsistent with at least one polynomial constraint, making conflicts unavoidable. A demonstration is provided in Example~\ref{ex:demo1}.

\begin{figure}
    \centering
    \includegraphics[scale=0.6]{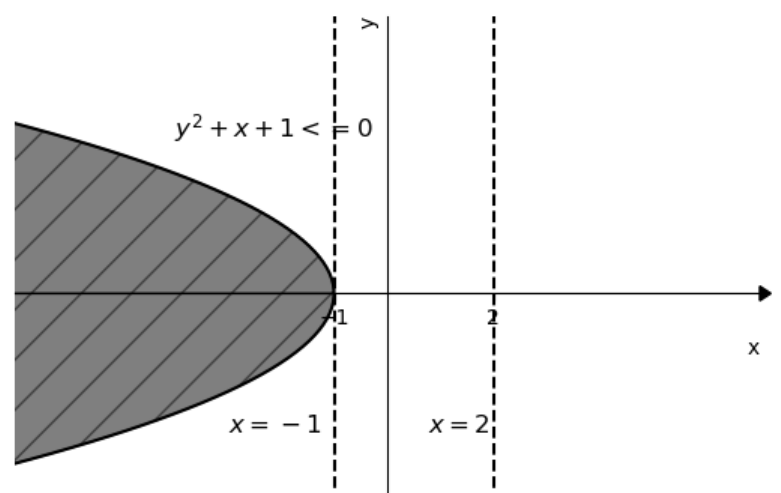}
    \caption{Demo of a conflict caused by a semantics decision.}
\label{fig:demo1}
\end{figure}

\begin{example}
\label{ex:demo1}
Consider the formula $y^2 + x + 1 \le 0$ with the variable order $\{x, y\}$. As depicted in Figure~\ref{fig:demo1}, if we decide $x \mapsto 2$, the satisfying region (shaded area) does not intersect the line $x = 2$, resulting in a conflict. This conflict is caused by an incorrect semantic decision for variable $x$ and could be avoided by choosing a correct value, for instance $x \mapsto -2$.
\end{example}

\subsection{Conflicts Caused by Literal Decisions}
A key technique in NLSAT is processing clauses that are univariate with respect to the current arithmetic variable. In a CDCL-style systematic search, literals are assigned either through unit propagation for unit clauses (i.e., clauses with only one unassigned literal) or via decisions for clauses with multiple unassigned literals. However, the literal decision mechanism in NLSAT has received relatively little attention. Improper literal decisions may introduce additional conflicts. We illustrate this situation in Example~\ref{ex:conflict1}.

\begin{example}
\label{ex:conflict1}
Consider the following three clauses:
\begin{align*}
c_1: y^2 + x - 2 \le 0 & \vee y^2 - x - 2 \le 0, \\
c_2: x + y = -3, & \quad c_3: x - y = 3.
\end{align*}

As illustrated in Figure~\ref{fig:demo1}, the purple area satisfies both polynomials in $c_1$, while the red and blue areas satisfy only $y^2 + x - 2 \le 0$ and $y^2 - x - 2 \le 0$, respectively. The straight lines represent the equality constraints in $c_2$ and $c_3$.

Suppose the SMT formula is $\{c_1, c_2\}$. The intersection occurs only in the red area. If the formula is $\{c_1, c_3\}$, the intersection is located only in the blue area. When NLSAT processes a clause with multiple unassigned literals such as $c_1$, it decides on one literal and branches the search space into either the red+purple or blue+purple areas. In this scenario, there is a 50\% chance of missing the equality line and encountering a conflict. 

However, the feasible-set of $c_1$ (i.e., the union of red, blue, and purple areas) has a nonempty intersection with both lines, indicating that both formulas $\{c_1, c_2\}$ and $\{c_1, c_3\}$ are indeed satisfiable.
\end{example}

\begin{figure}
    \centering
    \includegraphics[scale=0.6]{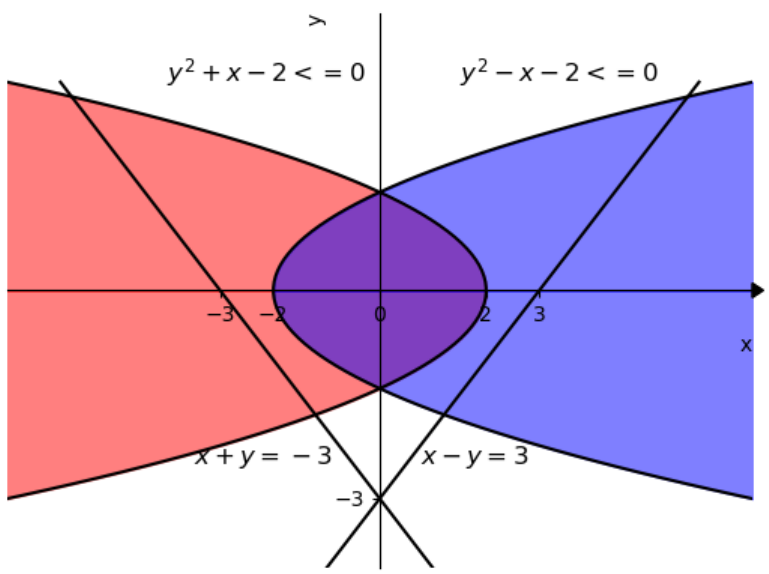}
    \caption{Demo of a conflict caused by a literal decision.}
\label{fig:demo1}
\end{figure}

Here comes a new problem in this circumstance. Is it possible to avoid conflicts with a better literal-decision heuristic? If so, how can this be achieved?
The key idea is to view the literal-decision problem as a satisfiability problem over intervals. Example~\ref{ex:conflict} illustrates this idea.

\begin{example}
\label{ex:conflict}
Suppose the current assignment is $\alpha := \{x \mapsto 0\}$. Clauses univariate with respect to $y$ are shown as follows:
\begin{align*}
c_1: \quad (y + 2)(y + 4) \le x \vee (y - 2)(y - 4) \le x \\
c_2: \quad (y + 5)(y + 6) \le x \vee (y - 1)(y - 5) \le x
\end{align*}
By calculating the feasible-set of each literal, the interval view of the clauses is:
\begin{flalign*}    
c_1: [-4, -2] \vee [2, 4] \\
c_2: [-6, -5] \vee [1, 5]
\end{flalign*}
Then the problem can be stated as: is there a value that belongs to at least one interval of each clause?
\end{example}

\section{Feasible-Set Based Look-Ahead Mechanism}
\label{sec:look-ahead}

In this section, we incorporate the clause-level feasible-set into the NLSAT algorithm and design a look-ahead mechanism.

\subsection{Look-Ahead Before Processing Clauses}

We first extend the definition of feasible-set to a set of clauses.

\begin{definition}
Given a clause set $CS$, an arithmetic variable $x$, and an assignment that maps all variables appearing in any clause of $CS$ except $x$, the \emph{feasible-set} (resp. \emph{infeasible-set}) is the set of values for $x$ that satisfy (resp. unsatisfy) all clauses in $CS$. 
Formally, the feasible-set of $CS$ can be computed as the intersection of the feasible-sets of individual clauses:
\[
\mathrm{feasible\_set}(CS) = \bigcap_{c \in CS} \mathrm{feasible\_set}(c)
\]
\end{definition}

By using the feasible-set of a clause set, we can more directly determine the search space of an arithmetic variable. Specifically, when the feasible-set is non-empty, assigning any value from the set to the variable (i.e., a semantics decision) guarantees progress in the search, allowing the algorithm to proceed to the next stage. Conversely, when the feasible-set is empty, no choice of value can avoid inconsistency. 

We now answer the question posed in Section~\ref{sec:conflict} with the following formal definition:

\begin{definition}
Given a clause set, if its feasible-set is non-empty, then, in theory, conflicts can be avoided by choosing appropriate literal assignments; we call this a \emph{path case}. Conversely, if the feasible-set is empty, conflicts cannot be avoided through literal decisions (caused by semantic decisions); we call this a \emph{block case}. Examples~\ref{ex:path} and \ref{ex:block} illustrate a path case and a block case, respectively.
\end{definition}

\begin{example}
\label{ex:path}
\begin{equation}
\left.
\begin{aligned}    
c_1: & [-4, -2] \vee \mybox[green]{\textnormal{[2, 4]}} \rightarrow \{[-4, -2] \cup [2, 4]\},\\
c_2: & [-6, -5] \vee \mybox[green]{\textnormal{[1, 5]}} \rightarrow \{[-6, -5] \cup [1, 5]\}
\end{aligned}
\right\} \bigwedge \rightarrow \{[2, 4]\}
\nonumber
\end{equation}
This example shows a path case, since the clauses can be satisfied by deciding literals in the green boxes.
\end{example}

\begin{example}
\label{ex:block}
\begin{equation}
\left.
\begin{aligned}    
c_1: & \mybox[red]{\textnormal{[-4, -2]}} \vee \mybox[red]{\textnormal{[2, 4]}} \rightarrow \{[-4, -2] \cup [2, 4]\},\\
c_2: & \mybox[red]{\textnormal{[-6, -5]}} \vee \mybox[red]{\textnormal{[5, 6]}} \rightarrow \{[-6, -5] \cup [5, 6]\} 
\end{aligned}
\right\} \bigwedge \rightarrow \varnothing
\nonumber
\end{equation}
This example shows a block case: the clauses cannot be satisfied regardless of which literals we decide.
\end{example}
 
The feasible-set computation provides a view of the currently consistent search space. However, in a CDCL-style algorithm, literals still need to be assigned to enable future conflict analysis. This raises the following question: how should we decide literals once we already know it is a path case (i.e., when the green boxes in Example~\ref{ex:path} are identified)? 

In our approach, we employ a look-ahead mechanism that first selects a pre-appointed value from the feasible-set. This pre-appointed value is then used to guide the search for a consistent decision path. The detailed procedure is presented in Algorithm~\ref{alg:process}.

\begin{algorithm}[!t]
\caption{Deciding Literals Using Pre-Appointed Value}
\label{alg:process}
\SetKwInOut{Input}{Input}
\SetKwInOut{Output}{Output}
\Input{A set of clauses $F$, pre-appointed value $val$ selected from feasible-set}
\Output{Decided literals $lits$}

$lits \gets \emptyset$\;

\For{each clause $c \in F$}{
    $path\_literal \gets$ \texttt{undefined}\;
    \For{each literal $l \in c$}{
        $lit\_set \gets$ compute feasible-set of $l$\;
        $val\_lit \gets$ real propagate literal $l$\;
        \If{$val\_lit = \top$}{
            break \tcp{clause already satisfied}
        }
        \If{$val\_lit = \bot$}{
            continue \tcp{literal unsatisfied, check next literal}
        }
        \If{$lit\_set$ contains $val$}{
            $path\_literal \gets l$ \tcp{decide satisfiable literal under pre-appointed value}
        }
    }
    \If{only one literal undefined in $c$}{
        unit propagate the literal\;
    }
    \ElseIf{$path\_literal$ is defined}{
        $lits \gets lits \cup \{path\_literal\}$\;
    }
}
\Return{$lits$}\;
\end{algorithm}

The updated algorithm introduces the additional condition that the feasible-set of the current literal must contain the pre-appointed value. This ensures that the feasible-sets of all decided literals during the processing procedure intersect at the clause-set level, allowing the arithmetic variable to be assigned the pre-appointed value. In the block case, the processing algorithm behaves identically to NLSAT, eventually triggering the resolve procedure to revise previous arithmetic assignments.

\subsection{Look-Ahead After Conflict Analysis}
In addition to processing clauses with multiple literals, our decision-making algorithm remains effective in conjunction with cylindrical algebraic decomposition (CAD)-based explanation. CAD projects conflict polynomials to learn a lemma that eliminates a sign-invariant cell. The learned lemma includes extended polynomial constraints, referred to as \emph{root atoms}, of the form:
$$
y \sim root_i(p(x_1, \dots, x_n)),
$$
where $y$ is the last assigned variable, $\sim \in \{ =, \neq, \le, \ge, <, > \}$, and $p$ is a polynomial generated via model-based projection, involving previously assigned variables $x_1, \dots, x_n$. Specifically, when the last assigned variable $y$ lies between two polynomial constraints, multiple root atoms may be generated in the learned lemma, making the choice of literal assignment particularly critical.

\begin{algorithm}[!t]
\caption{Process Clauses After a New Lemma}
\label{alg:lemma}
\SetKwInOut{Input}{Input}
\Input{A new lemma $lemma$, arithmetic variable $v$}

$\mathit{lemma\_feasible\_set} \gets$ compute feasible-set of clause($lemma$)\;
$\mathit{feasible\_set}[v] \gets \mathit{feasible\_set}[v] \cap \mathit{lemma\_feasible\_set}$\;

\If{$\mathit{feasible\_set}[v]$ is empty}{
    \tcp{Block case: no consistent assignment possible for $v$}
    \tcp{Proceed as in original NLSAT (Algorithm~\ref{alg:nlsat_process})}
    original process clauses\;
}
\Else{
    \tcp{Path case: a consistent assignment exists for $v$}
    $\mathit{val} \gets$ value\_selection($\mathit{feasible\_set}[v]$)\;
    \tcp{Call Algorithm~\ref{alg:process} to decide literals using the pre-appointed value}
    deciding literals using pre-appointed value ($\mathit{val}$)\;
}
\end{algorithm}

\begin{example}
\label{ex:lemma}
\begin{align*}    
c_1: & \mybox[green]{\textnormal{[-7, -2]}} \vee [2, 8] & \hspace{1em} c_1: [-7, -2] \vee \mybox[green]{\textnormal{[2, 8]}}\\
c_2: & [-11, -10] \vee \mybox[green]{\textnormal{[-6, 5]}} & \hspace{1em} c_2: [-11, -10] \vee \mybox[green]{\textnormal{[-6, 5]}} \\
learned: & \mybox[red]{\textnormal{[3, 4]}} \vee \mybox[red]{\textnormal{[7, 8]}} & \hspace{1em} learned: \mybox[green]{\textnormal{[3, 4]}} \vee [7, 8]
\end{align*}

Before learning a new lemma, the left column highlights a possible path in the green boxes. After incorporating the new lemma, the updated feasible-set may become inconsistent with any literal in the lemma, resulting in a conflict. The right column shows a new feasible path after incrementally processing the lemma.
\end{example}

Look-ahead algorithm after conflict analysis is similar to the main search part, as shown in Algorithm~\ref{alg:lemma}. The main difference is the incremental computation of decision cases by considering only the feasible-set of the new lemma. We use a vector of intervals to cache the previous feasible sets. After calculating the clause-level feasible-set, we must reprocess the clauses for the path case and find a new decision path. Because the newly generated lemma adds a new constraint on the arithmetic variable, the current decision path might be blocked as shown in Example~\ref{ex:lemma}.

\section{Clause-Level Propagation}
\label{sec:prop}

Following the idea of using a clause-level feasible-set, this section introduces a new kind of propagation called \emph{clause-level propagation}. The idea is inspired by unit propagation (or literal propagation) in SAT solving. In a boolean satisfaction problem, boolean variables can be unit propagated to assign a value, which allows the solver to detect conflicts as early as possible. However, most existing complete algorithms do not perform arithmetic propagation for quick assignment or conflict detection. 

We now formally define clause-level propagation.

\begin{definition}
\label{def:propagate}
Given a clause $c$, an arithmetic variable $x$, and an assignment $\alpha$ that assigns all variables appearing in $c$ except $x$, \emph{clause-level propagation} on $x$ is the computation of the feasible-set of the clause $c$, which serves to narrow the feasible-set of the variable $x$.
\end{definition}

The main difference between arithmetic and boolean problems lies in the structure of their search spaces. In SAT solving, unit propagation assigns boolean variables to either true or false. In other words, unit propagation always prunes the search space of a boolean variable by half, effectively guiding the search forward, since there are only two possible values. 

By contrast, in arithmetic problems, a clause may only eliminate part of the real-valued domain of a variable, contracting the search space without fully deciding the variable. In this section, we introduce the \emph{clause-level propagation} algorithm, which computes feasible-sets for arithmetic variables, and then show how to use this propagation information to guide the search by selecting the next branching variable.

\subsection{Clause-Level Propagation Method}
In SAT solving, unit propagation is performed after variable assignments. In our algorithm, we incrementally compute the feasible-set of a clause whenever it becomes univariate with respect to an arithmetic variable\footnote{This occurs not only after assigning an arithmetic variable, but also after assigning a boolean variable. Whenever a clause is arithmetically univariate, its feasible-set is updated.}. 

A newly generated univariate clause imposes an additional constraint on the arithmetic variable, pruning its search space by taking the intersection with the existing feasible-set. Unlike boolean search spaces, arithmetic search spaces may only be partially reduced. We categorize clause-level propagation into three cases, illustrated in Example~\ref{ex:propagation}:
\begin{itemize}
    \item \textbf{Block case:} The clause-level feasible-set is empty.
    \item \textbf{Fixed case:} The clause-level feasible-set contains exactly one real number, e.g., $[2, 2]$.
    \item \textbf{Other case:} The clause-level feasible-set is narrowed but neither empty nor a single value.
\end{itemize}

This propagation information is subsequently used to guide the branching heuristic. The algorithmic details are shown in Algorithm~\ref{alg:propagation}.

\begin{algorithm}[!t]
\caption{Clause-Level Propagation}
\label{alg:propagation}
\SetKwInOut{Input}{Input}
\Input{Clause set $F$, current feasible sets of arithmetic variables}

\For{each clause $\mathit{cls} \in F$}{
    \If{$\mathit{cls}$ is univariate to an arithmetic variable $v$}{
        $\mathit{cls\_feasible\_set} \gets$ compute\_feasible\_set($\mathit{cls}$)\;
        $\mathit{feasible\_set[v]} \gets \mathit{feasible\_set[v]} \cap \mathit{cls\_feasible\_set}$\;

        \tcp{Categorize propagation result}
        \If{$\mathit{feasible\_set[v]}$ is empty}{
            $\mathit{blocked\_vars} \gets \mathit{blocked\_vars} \cup \{v\}$ 
            \tcp*{Block case: conflict unavoidable}
        }
        \ElseIf{$\mathit{feasible\_set[v]}$ is a single value}{
            $\mathit{fixed\_vars} \gets \mathit{fixed\_vars} \cup \{v\}$ \tcp*{Fixed case: value determined}
        }
        \Else{
            \tcp{Other case: search space narrowed but not fixed}
        }
    }
}
\end{algorithm}

\begin{example}
\label{ex:propagation}
An example is shown in Figure~\ref{fig:propagation}.  
When a variable $x$ is assigned $0$, three clauses become univariate to other variables $\{z, y, k\}$.  
These three clauses add three new constraints on arithmetic variables, calculated as
\[
\{(-\infty, -2] \cup [2, 6]\}, \quad \{[2, 2]\}, \quad \emptyset.
\]  
These feasible sets indicate that variables are \textbf{feasible}, \textbf{fixed}, or \textbf{blocked}.
\begin{figure}[h]
    \centering
    \includegraphics[width=0.45\textwidth]{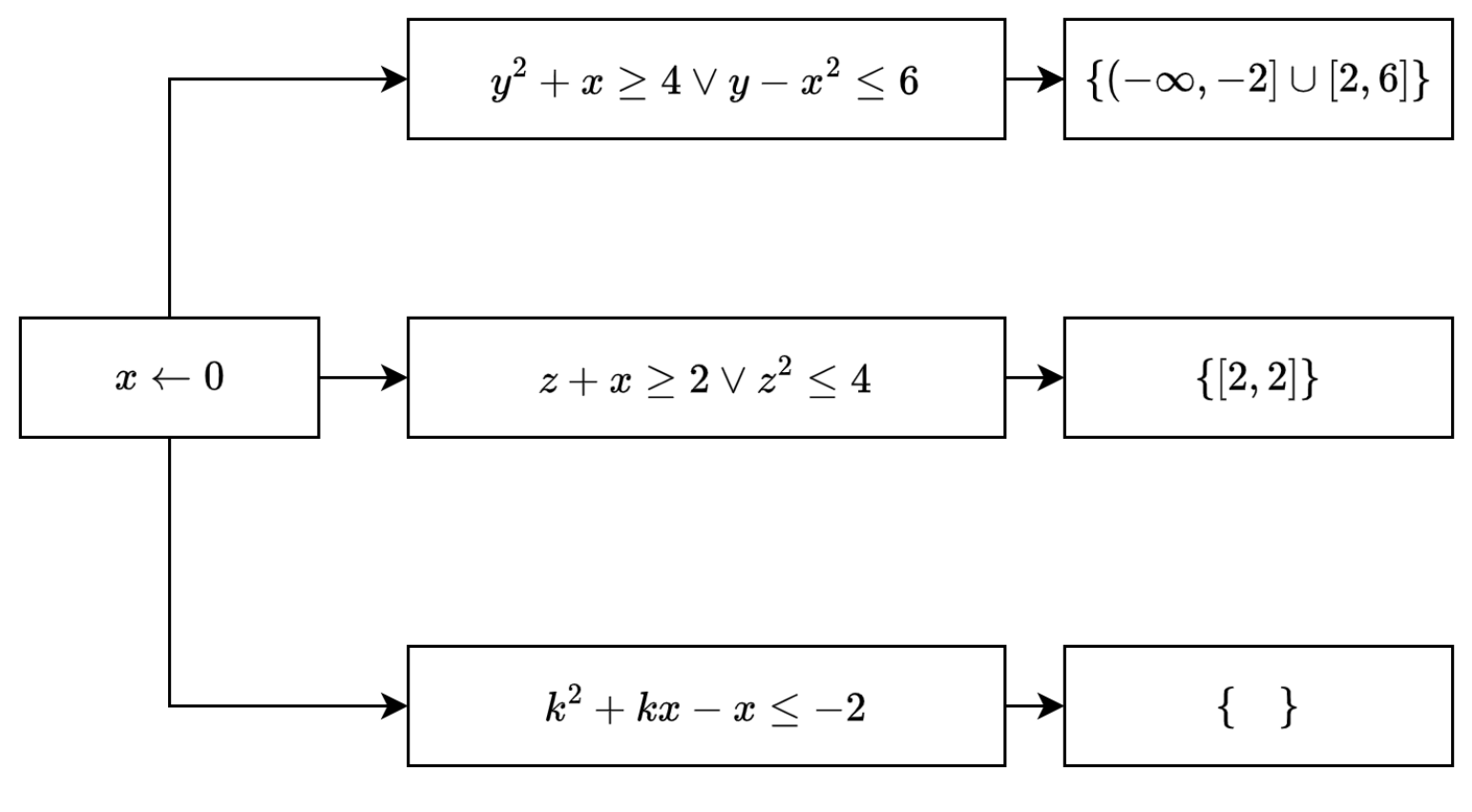}
    \caption{Demo of clause-level propagation for $y$ (\textbf{normal case}), $z$ (\textbf{fixed case}) and $k$ (\textbf{block case}).}
    \label{fig:propagation}
\end{figure}
\end{example}

\subsection{Propagation-Based Branching Heuristic} 
In SAT solving, after unit propagation, an unassigned boolean variable is either propagated to a value or a conflict clause is detected immediately. Similarly, for arithmetic variables, the clause-level feasible-set allows us to identify propagation and conflict cases, corresponding to the \emph{fixed} and \emph{blocked} cases introduced above. 

However, unlike boolean unit propagation, clause-level conflicts for arithmetic variables cannot directly return a conflict clause in NLSAT; this task remains the responsibility of the clause processing procedure. Therefore, we record information about fixed and blocked variables and prioritize them in the branching heuristic, ensuring that the search addresses these critical variables as early as possible. 

For variables in the normal case (neither fixed nor blocked), we adopt the Variable State Independent Decaying Sum (VSIDS) heuristic~\cite{vsids}, as suggested in~\cite{MCSAT2, MCSATOrder}. The full procedure is outlined in Algorithm~\ref{alg:branching}. Details on our implementation of dynamic variable ordering are discussed in Section~\ref{sec:imp}.

\begin{algorithm}[!t]
\caption{Propagation-Based Branching Heuristic}
\label{alg:branching}
\SetKwInOut{Output}{Output}
\Output{A variable $v$ to branch on}
\textcolor{red}{
\tcp{Prioritize blocked variables (potential conflicts) first}
\If{$\mathit{blocked\_vars} \neq \emptyset$}{
    $v \gets \text{select\_from}(\mathit{blocked\_vars})$\;
}}

\textcolor{blue}{
\tcp{Next, consider fixed variables (propagate value)}
\ElseIf{$\mathit{fixed\_vars} \neq \emptyset$}{
    $v \gets \text{select\_from}(\mathit{fixed\_vars})$\;
}}
\tcp{Otherwise, use VSIDS heuristic for normal variables}
\Else{
    $v \gets \text{vsids\_select}()$\;
}
\Return $v$
\end{algorithm}

\section{Implementation}
\label{sec:imp}

\begin{algorithm}[!t]
\caption{ClauseSMT}
\label{alg:clausesmt}
\SetKwInOut{Input}{Input}
\SetKwInOut{Output}{Output}

\Input{A formula $F$}
\Output{SAT or UNSAT}

\While{true}{
    clause-level propagation($F$)\; \tcp{call Algorithm~\ref{alg:propagation}}
    variable $v \gets$ branching heuristic\; \tcp{call Algorithm~\ref{alg:branching}}
    
    \If{$v$'s feasible-set is empty}{
        $new\_lemma \gets$ Resolve (Conflict Analysis)\;
        \If{$new\_lemma$ is empty}{
            \Return UNSAT\;
        }
        \Else{
            Process Clauses after a new lemma($new\_lemma$)\; \tcp{call Algorithm~\ref{alg:lemma}}
        }
    }
    \Else{
        $val \gets$ select from feasible-set\;
        Process Clauses using pre-appointed value $val$\; \tcp{call Algorithm~\ref{alg:process}}
        assign $v \gets val$\;
        \If{all variables are assigned}{
            \Return SAT\;
        }
    }
}
\end{algorithm}

\begin{figure}
    \centering
    \includegraphics[width=0.45\textwidth]{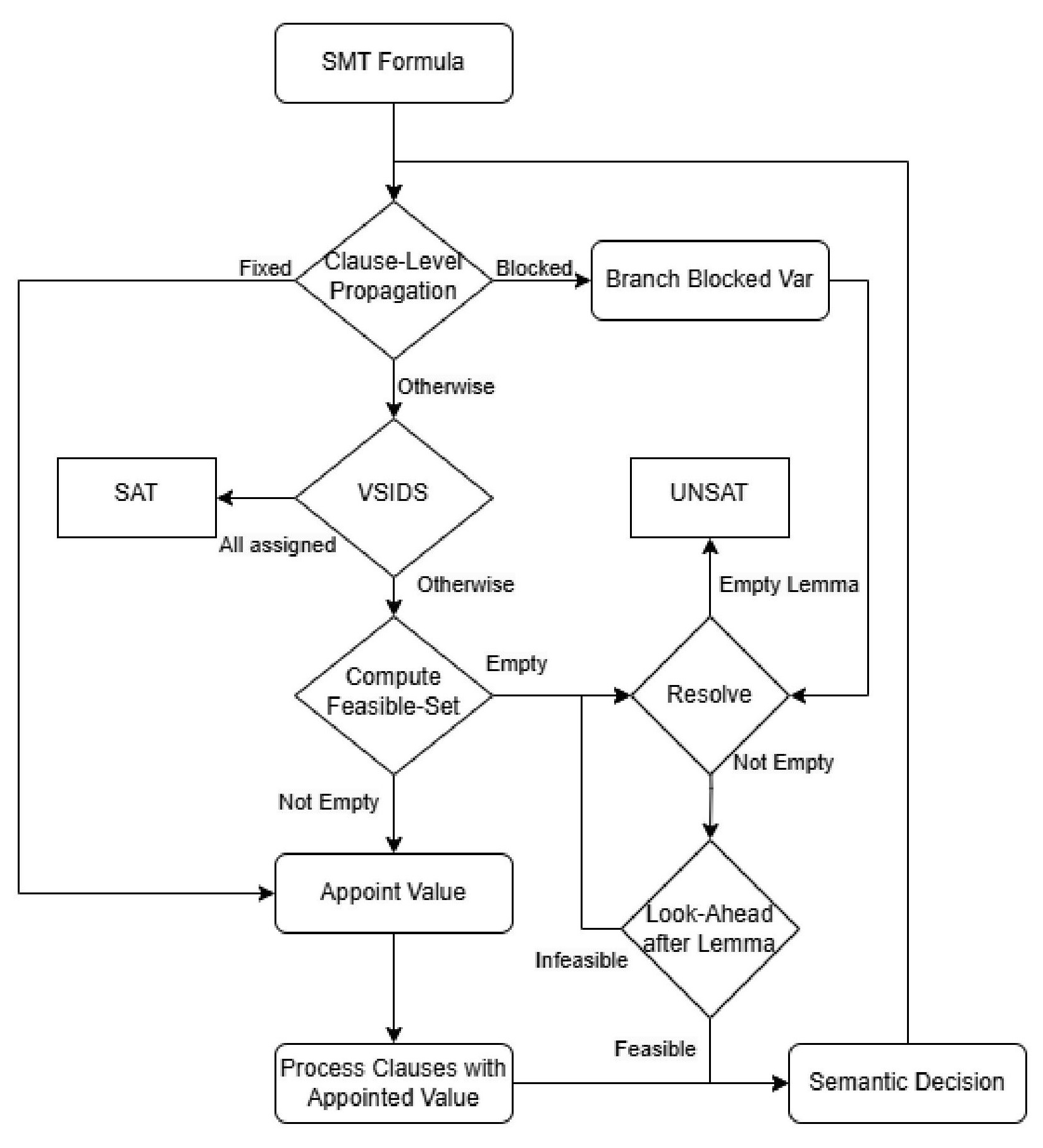}
    \caption{Overall Structure of clauseSMT.}
    \label{fig:structure}
\end{figure}

All of the above algorithms are incorporated into our new solver, \texttt{clauseSMT}. We describe the solver in detail, including an extended version of conflict analysis and an implementation of the dynamic variable ordering framework. The overall structure of \texttt{clauseSMT} is illustrated in Figure~\ref{fig:structure}, and Algorithm~\ref{alg:clausesmt} presents the integrated heuristics implemented in the solver.

\subsection{Resolve}
Although the look-ahead based processing algorithm is introduced in the previous section, practical implementation involves more complex cases. As discussed, conflicts in NLSAT can be categorized into two types:

\begin{enumerate}
    \item \textbf{Literal-decision conflicts}~\footnote{For the look-ahead mechanism, this occurs only in block cases.} \\
    For these, we keep the resolve algorithm the same as in the original NLSAT. The search engine backtracks to the highest decision level of the new lemma and attempts to unit propagate the negation of previously decided literals. In other words, this type of conflict occurs purely at the literal level.

    \item \textbf{Incremental clause conflicts} \\
    For literal-decision conflicts considering an incremental clause (i.e., a newly learned lemma), we detect these by recalculating the decision path. Any literals that were decided prior to this stage must be reset to allow proper propagation according to the updated feasible sets.
\end{enumerate}

To better manage the backtracking process, we introduce several new types of trails, commonly used in the NLSAT algorithm:

\begin{itemize}
    \item \textbf{path\_finder}: Whenever the feasible-set at the current stage is non-empty, a path exists and this trail is recorded.
    \item \textbf{block\_finder}: When the current stage is blocked, this trail is recorded.
    \item \textbf{clause\_feasible\_updated}: Whenever the feasible-set of a clause for an arithmetic variable is updated, this trail is recorded.
\end{itemize}

\subsection{Dynamic Variable Ordering Framework}
\subsubsection{Watched Variables}

We implement two \emph{watched variables}, inspired by the two-watched-literals scheme, to detect univariate clauses and lemmas. Each clause or lemma is watched by two variables (boolean or arithmetic) appearing in it. Watchers are updated whenever one of them is assigned. The cases are as follows:

\begin{itemize}
    \item There exists a third variable unassigned: we replace the assigned watcher with this variable.
    \item No other variable is unassigned: this clause is univariate to the remaining unassigned watcher.
    \item Both watchers are assigned: no action is taken.
\end{itemize}

Whenever a univariate clause is detected, its feasible-set is updated eagerly, which helps the search engine gather more clause-level information.

\subsubsection{Projection Order}

A common challenge in nonlinear arithmetic is the strict variable-order relationship of root atoms, which are generated by model-based projection. Given a polynomial set $ps$ and a projection order $\{v_1, v_2, \dots, v_k\}$, each time the projection method eliminates a variable, it generates a root atom corresponding to that variable. 

As discussed in~\cite{MCSATOrder}, in most cases, the variable not appearing in the polynomial should be assigned last. Specifically, when all atoms are in root format, the projection order should exactly be the inverse of the assignment order, which is how it is implemented in our solver.

\subsubsection{Branching Heuristic}
VSIDS is a particularly effective branching heuristic. Each time a conflict is detected, we increase the activities of the involved variables of all types. Following the design in~\cite{MCSATOrder}, we employ several branching heuristics and use the \textbf{uniform} heuristic to compare the activity of boolean and arithmetic variables on the same scale.

\subsubsection{Parameter Settings}
Values of the tunable parameters are summarized in Table~\ref{tab:parameters}.

\begin{table}[H]
\small
\centering
\begin{tabularx}{0.95\linewidth}{p{2.25cm}|X|c}
  \toprule
  Symbol & Description & Value \\
  \midrule
  $\mathit{arith\_decay}$ & Decay factor for arithmetic variables & 0.95 \\
  $\mathit{bool\_decay}$ & Decay factor for boolean variables & 0.95 \\
  $\mathit{arith\_bump}$ & Incremental amount of arithmetic activity & 1 \\
  $\mathit{bool\_bump}$ & Incremental amount of boolean activity & 1 \\
  $\mathit{lemma\_conf}$ & Initial conflict count for deleting lemmas & 100 \\
  $\mathit{lemma\_conf\_inc}$ & Incremental factor for lemma conflicts & 1.5 \\
  \bottomrule
\end{tabularx}
\vspace{1mm}
\caption{Tunable parameters}
\label{tab:parameters}
\end{table}

\subsection{Shortcut for UNSAT Instances}
For the block case discussed in Section~\ref{sec:look-ahead}, we process blocked clauses the same way as in NLSAT. As concluded earlier, conflicts in this scenario occur because previous variables (arithmetic or boolean) were assigned incorrect values. In our implementation, we introduce a shortcut mechanism to directly return \texttt{UNSAT} if the blocked clauses involve only a single variable. In this situation, there are no previous stages, and thus the instance is guaranteed to be unsatisfiable.

\section{Evaluation}
\label{sec:eva}
In this section, we compare our algorithm with several existing solvers, including Z3 (version 4.13.1)~\cite{MouraB08}, CVC5 (version 1.0.2)~\cite{BarbosaBBKLMMMN22}, and YICES2 (version 2.6.2)~\cite{Dutertre14}. We also present an ablation study to analyze the impact of various improvements.

\subsection{Experiment Preliminaries}
The standard benchmark for evaluating SMT solvers is SMT-LIB\footnote{\url{https://smt-lib.org/}}. The full benchmark for the QF\_NRA theory consists of 12,134 instances, originating from various applications, including nonlinear hybrid automata, ranking function generation for program analysis, and other mathematical problems. Most instances are labeled as \texttt{SAT} or \texttt{UNSAT}, though some remain \texttt{UNKNOWN}. It should be noted that instances from SMT-LIB exhibit significant variation in clause numbers, literal counts, and polynomial degrees. Our experiments are conducted on a server equipped with an Intel Xeon Platinum 8153 processor running at 2.00 GHz. Each instance is limited to a maximum runtime of 1,200 seconds, consistent with the SMT-COMP settings.

\subsection{RQ1: Comparison with mainstream Solvers}
We compare our algorithm with other SMT solvers in Table~\ref{tab:instances}. When evaluating our approach using Z3, we disable all other tactics such as CDCL(T) and any incomplete algorithms. The solvers Z3, CVC5, and YICES2 are tested without modifications, each employing its portfolio of different algorithms. We also evaluate the original NLSAT solver by disabling all other tactics.

\begin{table}[!t]
\centering
\tiny
\setlength{\tabcolsep}{4pt}
\begin{tabular}{
  >{\centering\arraybackslash}p{2.5cm}
  | c | c | c | c | c | c | c
}
\hline
Category & \multicolumn{2}{c}{\#inst} & Z3 & YICES2 & CVC5 & NLSAT & Ours \\\hline
\multirow{2}*[-1.5ex]{20161105-Sturm-MBO} & \multirow{2}*[-1.5ex]{405} 
& SAT & 0 & 0 & 0 & 0 & 0 \\\cline{3-8}
& & UNSAT & 124 & \textbf{285} & \textbf{285} & 44 & 39 \\\cline{3-8}
& & SOLVED & 124 & \textbf{285} & \textbf{285} & 44 & 39 \\\hline
\multirow{2}*[-1.5ex]{20161105-Sturm-MGC} & \multirow{2}*[-1.5ex]{9} 
& SAT & \textbf{2} & 0 & 0 & \textbf{2} & \textbf{2} \\\cline{3-8}
& & UNSAT & \textbf{7} & 0 & 0 & \textbf{7} & 6 \\\cline{3-8}
& & SOLVED & \textbf{9} & 0 & 0 & \textbf{9} & 8 \\\hline
\multirow{2}*[-1.5ex]{20170501-Heizmann} & \multirow{2}*[-1.5ex]{69} 
     & SAT & \textbf{2} & 0 & 1 & 1 & \textbf{2} \\\cline{3-8}
 & & UNSAT & 1 & 12 & 9 & 10 & \textbf{19} \\\cline{3-8}
& & SOLVED & 3 & 12 & 10 & 11 & \textbf{21} \\\hline
\multirow{2}*[-1.5ex]{20180501-Economics-Mulligan} & \multirow{2}*[-1.5ex]{135} 
     & SAT & \textbf{93} & 91 & 89 &\textbf{93} & 92 \\\cline{3-8}
 & & UNSAT & 39 & 39 & 35 & \textbf{41} & \textbf{41} \\\cline{3-8}
& & SOLVED & 132 & 130 & 124 & \textbf{134} & 131 \\\hline
\multirow{2}*[-1.5ex]{2019-ezsmt} & \multirow{2}*[-1.5ex]{63} 
     & SAT & 56 & 52 & 50 & \textbf{58} & 36 \\\cline{3-8}
 & & UNSAT & \textbf{2} & \textbf{2}& \textbf{2} & \textbf{2} & \textbf{2} \\\cline{3-8}
& & SOLVED & 58 & 54 & 52 & \textbf{60} & 38 \\\hline
\multirow{2}*[-1.5ex]{20200911-Pine} & \multirow{2}*[-1.5ex]{245} 
     & SAT & 234 & \textbf{235} & 199 & \textbf{235} & 234 \\\cline{3-8}
 & & UNSAT & 6 & \textbf{8} & 5 & 7 & 5 \\\cline{3-8}
& & SOLVED & 240 & \textbf{243} & 204 & 242 & 239 \\\hline
\multirow{2}*[-1.5ex]{20211101-Geogebra} & \multirow{2}*[-1.5ex]{112} 
     & SAT & \textbf{110} & 99 & 91 & \textbf{110} & 98 \\\cline{3-8}
 & & UNSAT & 0 & 0 & 0 & 0 & 0 \\\cline{3-8}
& & SOLVED & \textbf{110} & 99 & 91 & \textbf{110} & 98 \\\hline
\multirow{2}*[-1.5ex]{20220314-Uncu} & \multirow{2}*[-1.5ex]{225} 
     & SAT & 69 & \textbf{70} & 62 & 68 & \textbf{70} \\\cline{3-8}
 & & UNSAT & \textbf{155} & 153 & 148 & \textbf{155} & 152 \\\cline{3-8}
& & SOLVED & \textbf{224} & 223 & 210 & 223 & 222 \\\hline
\multirow{2}*[-1.5ex]{hong} & \multirow{2}*[-1.5ex]{20} 
     & SAT & 0 & 0 & 0 & 0 & 0 \\\cline{3-8}
 & & UNSAT & 8 & \textbf{20} & \textbf{20} & 12 & 14 \\\cline{3-8}
& & SOLVED & 8 & \textbf{20} & \textbf{20} & 12 & 14 \\\hline
\multirow{2}*[-1.5ex]{hycomp} & \multirow{2}*[-1.5ex]{2752} 
     & SAT & \textbf{307} & 227 & 225 & 244 & 291 \\\cline{3-8}
 & & UNSAT & \textbf{2242} & 2201 & 2212 & 2088 & 2181 \\\cline{3-8}
& & SOLVED & \textbf{2549} & 2428 & 2437 & 2332 & 2472 \\\hline
\multirow{2}*[-1.5ex]{kissing} & \multirow{2}*[-1.5ex]{45} 
     & SAT & \textbf{33} & 10 & 17 & 12 & 14 \\\cline{3-8}
 & & UNSAT & 0 & 0 & 0 & 0 & 0 \\\cline{3-8}
& & SOLVED & \textbf{33} & 10 & 17 & 12 & 14 \\\hline
\multirow{2}*[-1.5ex]{LassoRanker} & \multirow{2}*[-1.5ex]{821} 
     & SAT & 167 & 122 & \textbf{305} & 220 & 302 \\\cline{3-8}
 & & UNSAT & 151 & 260 & \textbf{470} & 174 & 311 \\\cline{3-8}
& & SOLVED & 318 & 382 & \textbf{775} & 394 & 613 \\\hline
\multirow{2}*[-1.5ex]{meti-tarski} & \multirow{2}*[-1.5ex]{7006} 
     & SAT & \textbf{4391} & 4369 & 4343 & \textbf{4391} & 4372 \\\cline{3-8}
 & & UNSAT & 2605 & 2588 & 2581 & \textbf{2611} & 2588 \\\cline{3-8}
& & SOLVED & 6996 & 6957 & 6924 & \textbf{7002} & 6960 \\\hline
\multirow{2}*[-1.5ex]{UltimateAutomizer} & \multirow{2}*[-1.5ex]{61} 
     & SAT & 35 & 39 & 35 & \textbf{45} & 39 \\\cline{3-8}
 & & UNSAT & 11 & 12 & 10 & \textbf{13} & 12 \\\cline{3-8}
& & SOLVED & 46 & 51 & 45 & \textbf{58} & 51 \\\hline
\multirow{2}*[-1.5ex]{zankl} & \multirow{2}*[-1.5ex]{166} 
     & SAT & \textbf{70} & 58 & 58 & 62 & 56 \\\cline{3-8}
 & & UNSAT & 28 & \textbf{32} & \textbf{32} & 27 & 30 \\\cline{3-8}
& & SOLVED & \textbf{98} & 90 & 90 & 89 & 86 \\\hline
\multirow{2}*[-1.5ex]{Total} & \multirow{2}*[-1.5ex]{12134} 
     & SAT & 5569 & 5372 & 5475 & 5541 & \textbf{5608} \\\cline{3-8}
 & & UNSAT & 5379 & 5612 & \textbf{5809} & 5191 & 5397 \\\cline{3-8}
& & SOLVED & 10948 & 10984 & \textbf{11284} & 10732 & 11005 \\\hline
\end{tabular}
\vspace{2mm}
\caption{Summary of results for all instances in SMT-LIB (QF\_NRA).}
\label{tab:instances}
\end{table}
Our algorithm demonstrates competitive performance compared to state-of-the-art solvers such as Z3 and CVC5. In particular, clauseSMT solves the largest number of satisfiable instances and ranks third for unsatisfiable ones. 

\begin{figure}[!t]
     \centering
      \includegraphics[width=0.2\textwidth]{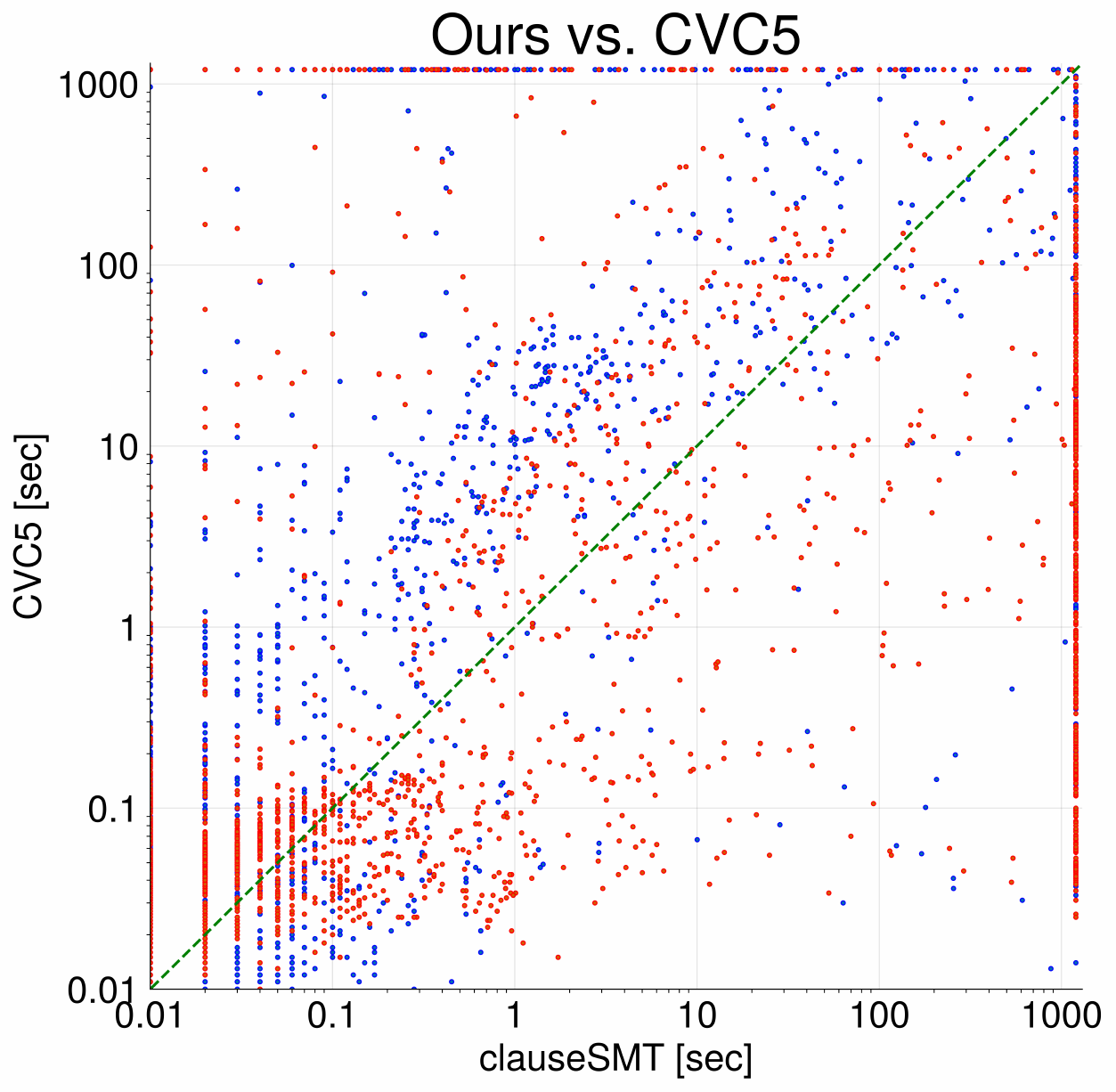}\qquad
    \includegraphics[width=0.2\textwidth]{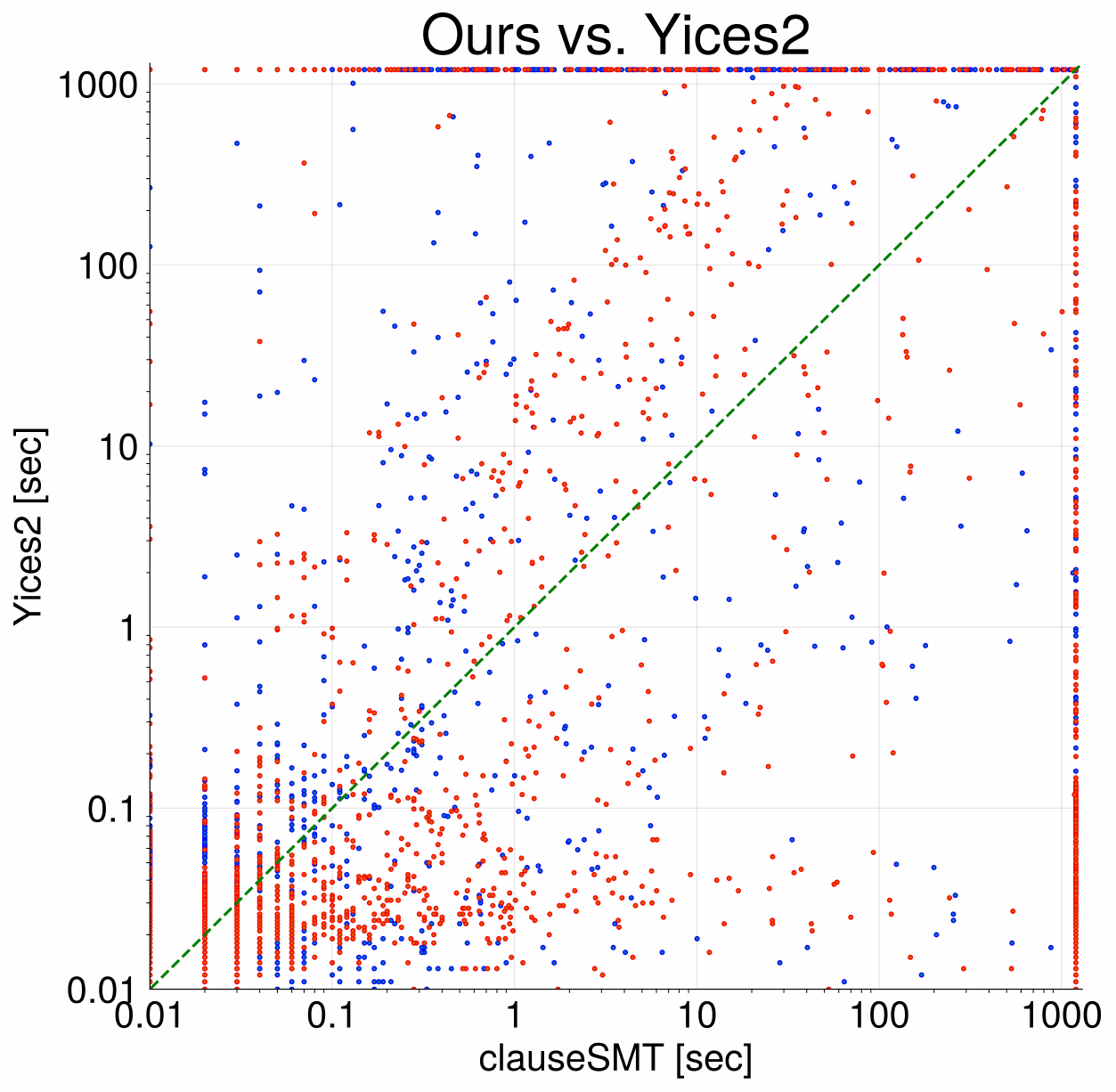}\qquad
     \includegraphics[width=0.2\textwidth]{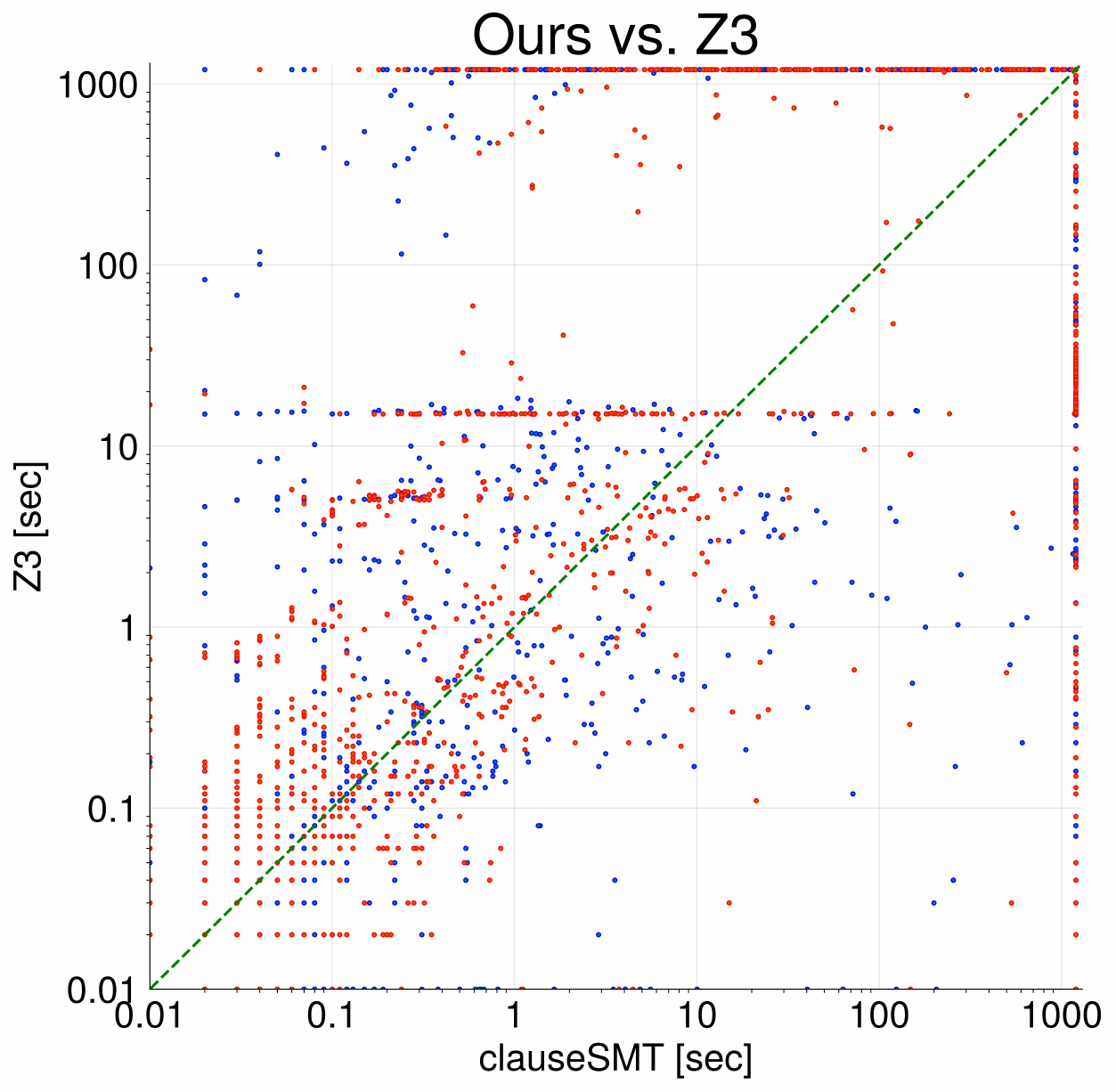}\qquad
    \includegraphics[width=0.2\textwidth]{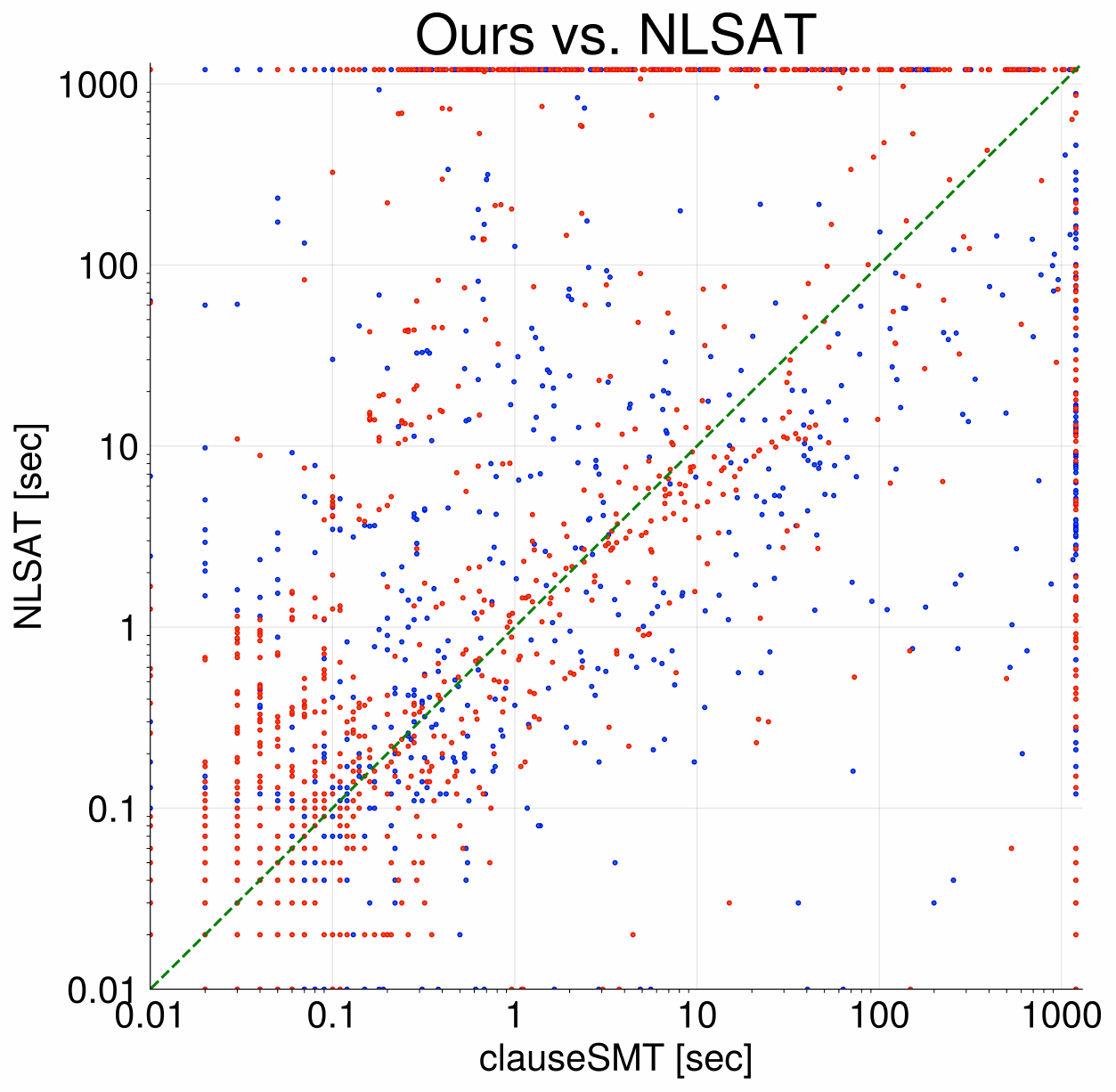}
        \caption{Run time comparison against CVC5, Z3, YICES2 and nlsat (blue points: satisfiable instances, red points: unsatisfiable instances).}
\label{fig:scatter}
\end{figure}

Figure~\ref{fig:scatter} presents pairwise scatter plots of solving times, where each point's x-coordinate corresponds to clauseSMT and the y-coordinate to a competing solver. Points below the diagonal ($y=x$) indicate instances where clauseSMT is faster. 

The plots show that clauseSMT efficiently handles satisfiable instances (blue points), often outperforming mainstream solvers, consistent with the results in Table~\ref{tab:instances}. The comparison with the original NLSAT baseline further highlights the benefits of our look-ahead and arithmetic-aware propagation techniques, which reduce solving time on most instances.

\subsubsection{Comparison with CVC5}
CVC5 solves the largest number of instances overall, particularly excelling in unsatisfiable cases due to incomplete techniques such as interval constraint propagation and incremental linearization. Notably, the MBO category~\cite{AkutsuHT08}, which contains single clauses with very high degrees, remains challenging for CAD-based algorithms.

\subsubsection{Comparison with Original NLSAT}
Our solver performs slightly worse on instances with high-degree polynomials, where feasible-set computations are relatively costly. Nevertheless, it demonstrates significant gains in LassoRanker and Hycomp, which contain thousands of instances with complex feasible-set relationships. For LassoRanker, our solver improves solved instances by 50\%, and overall it solves almost 300 more instances compared to the original NLSAT.

\subsection{RQ2: Effectiveness of Look-Ahead Mechanism}
To evaluate the impact of the look-ahead mechanism, we implement several variants summarized in Table~\ref{tab:effect_look-ahead}:
\begin{itemize}
    \item \textbf{Look-Ahead}: Feasible-set based look-ahead on original NLSAT with static variable order.
    \item \textbf{Lower Degree}: Decide literals with the lowest polynomial degree (default NLSAT heuristic).
    \item \textbf{Random Decide}: Randomly select literals when processing clauses.
\end{itemize}

Although the difference in solved instances across most categories is small, our algorithm solves about 50 more instances in the Hycomp category~\cite{CimattiMT12}, which contains numerous nonlinear equalities. These instances often exhibit literal path cases, highlighting the advantage of the look-ahead mechanism.

\begin{figure}
     \centering
     \includegraphics[width=0.2\textwidth]{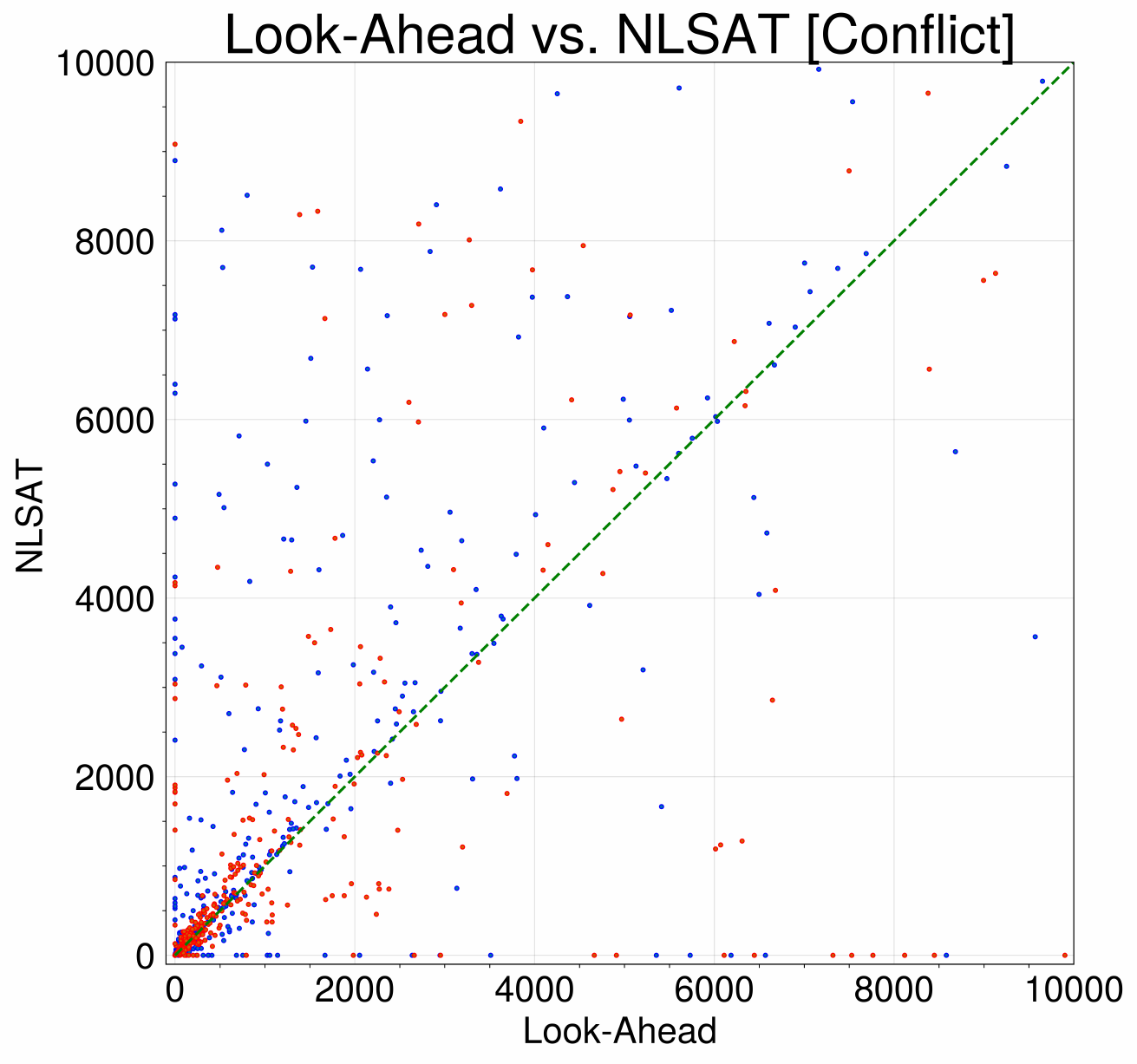}\qquad
     \includegraphics[width=0.2\textwidth]{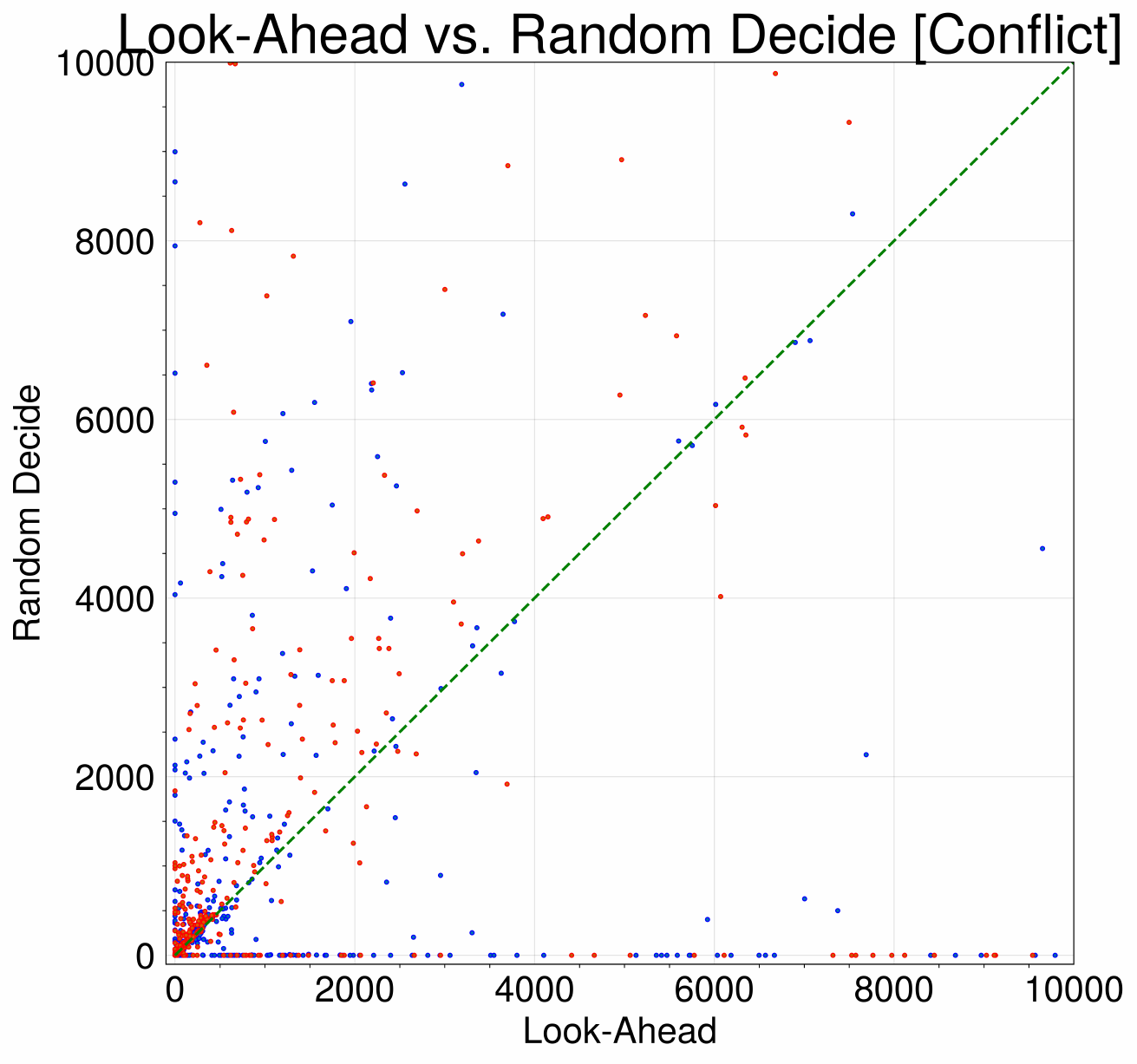}
    \caption{Conflict counts of look-ahead NLSAT versus original and random NLSAT (blue: satisfiable, red: unsatisfiable).}
\label{fig:look-ahead-conflict}
\end{figure}

The look-ahead mechanism proactively detects blocking cases and mitigates conflicts during path exploration. Figure~\ref{fig:look-ahead-conflict} presents a scatter plot comparing the number of conflicts incurred by the two algorithms, with the green line indicating parity. For most instances, the look-ahead strategy reduces conflicts, which—although not substantially impacting the 1200-second timeout—enables a more efficient systematic search, particularly when handling clauses with multiple literals. Interestingly, a few instances exhibit increased conflicts under the look-ahead mechanism. This arises from differences in sampling intervals used to select arithmetic assignments: the look-ahead algorithm employs the intersected interval for witness selection, whereas the baseline relies on the literal interval. Consequently, even with identical random seeds, the arithmetic variables may be assigned different values, leading to divergent subsequent search processes.

\begin{table}[!t]
\centering
\resizebox{\linewidth}{!}{%
\begin{tabular}{c | c | c | c | c}
Category & \#inst & Decide Lower Degree & Random Decide & Look-Ahead \\\hline
20161105-Sturm-MBO & 405 & 44 & 45 & 44 \\
20161105-Sturm-MGC & 9 & 9 & 9 & 9 \\
20170501-Heizmann & 69 & 11 & 5 & 7 \\
20180501-Economics-Mulligan & 135 & 134 & 134 & 134 \\
2019-ezsmt & 63 & 60 & 59 & 58 \\
20200911-Pine & 245 & 242 & 242 & 243 \\
20211101-Geogebra & 112 & 110 & 109 & 110 \\
20220314-Uncu & 225 & 223 & 224 & 224 \\
hong & 20 & 12 & 12 & 12 \\
hycomp & 2752 & 2332 & 2272 & 2388 \\
kissing & 45 & 12 & 14 & 15 \\
LassoRanker & 821 & 394 & 393 & 389 \\
meti-tarski & 7006 & 7002 & 7001 & 7002 \\
UltimateAutomizer & 61 & 58 & 44 & 57 \\
zankl & 166 & 89 & 89 & 87 \\\hline
Total & 12134 & 10732 & 10652 & 10778\\
\end{tabular}%
}
\vspace{1mm}
\caption{Comparison of solved instances for different literal decision mechanisms.}
\label{tab:effect_look-ahead}
\end{table}

\subsection{RQ3: Effectiveness of Clause-Level Propagation}
To evaluate clause-level propagation, we compare three solver versions: (1) original NLSAT with static variable order based on degree (\emph{static}), (2) dynamic NLSAT with VSIDS (\emph{VSIDS}), and (3) dynamic NLSAT with clause-level propagation (\emph{prop-VSIDS}). Results are summarized in Table~\ref{tab:effect_prop}.

\begin{figure}
     \centering
     \small
    \includegraphics[scale=0.2]{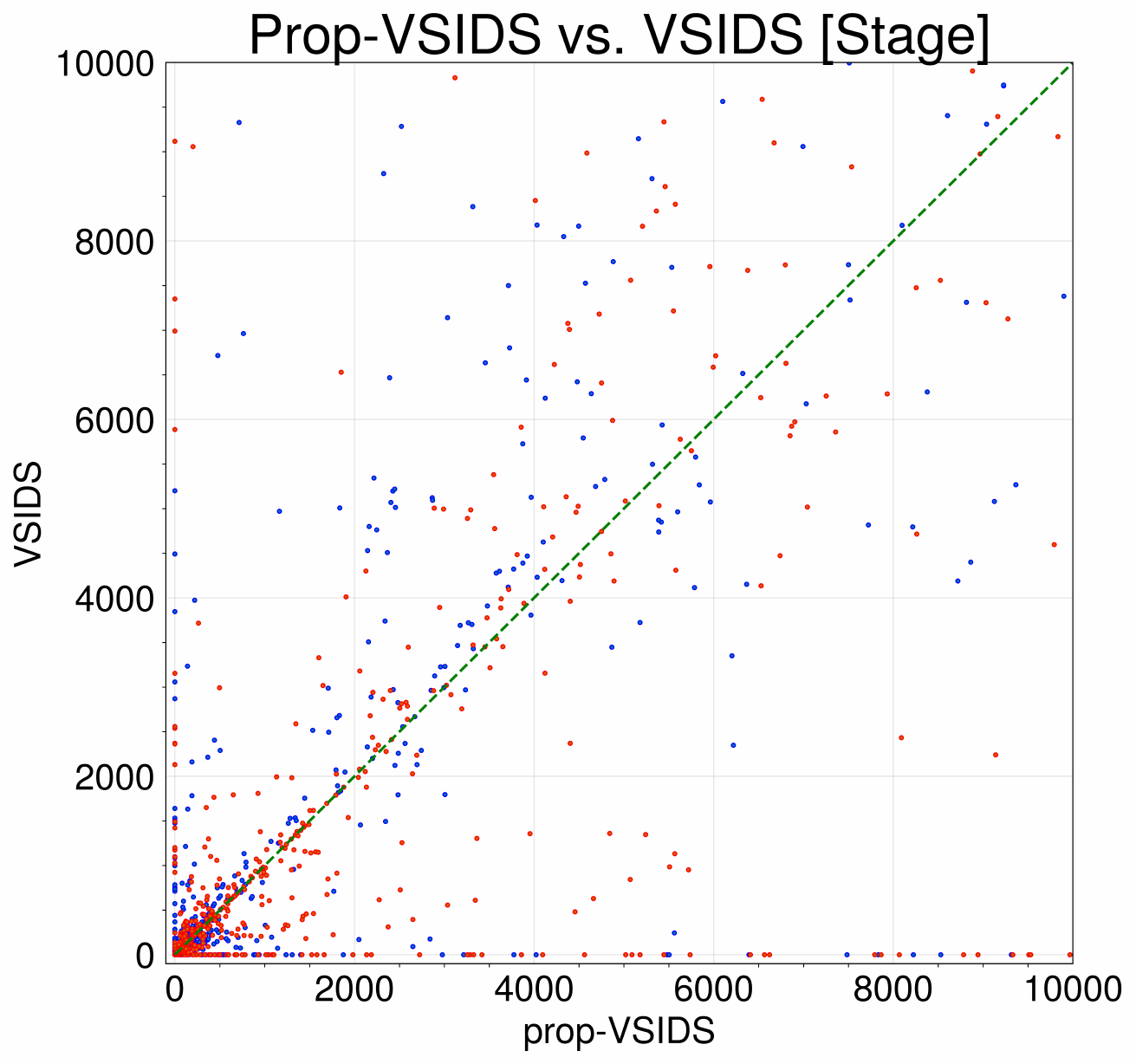}
        \caption{Stage comparison of prop-VSIDS against VSIDS (blue points: satisfiable instances, red points: unsatisfiable instances).}
\label{fig:prop-stage}
\end{figure}

The data show that VSIDS significantly improves performance within the MCSAT framework. Incorporating clause-level propagation further accelerates conflict detection and increases the number of solved instances across most categories. In particular, \emph{prop-VSIDS} excels on \texttt{hycomp}~\cite{CimattiMT12}, \texttt{LassoRanker}~\cite{HeizmannHLP13, LeikeH15}, and \texttt{meti-tarski}~\cite{AkbarpourP10}, all of which contain numerous arithmetic clauses prone to block cases. Figure~\ref{fig:prop-stage} illustrates the reduction in search stages (semantic decision steps) for \emph{prop-VSIDS} compared to traditional VSIDS, showing that inconsistent branching choices are detected earlier and overall stages are significantly reduced.

\begin{table}[!t]
\centering
\small
\resizebox{0.95\linewidth}{!}{%
\begin{tabular}{>{\centering\arraybackslash}p{4cm} | c | c | c | c}
Category & \#inst & Static & VSIDS & prop-VSIDS \\\hline
20161105-Sturm-MBO & 405 & 44 & 38 & 39 \\
20161105-Sturm-MGC & 9 & 9 & 6 & 8 \\
20170501-Heizmann & 69 & 7 & 20 & 21 \\
20180501-Economics-Mulligan & 135 & 134 & 133 & 133 \\
2019-ezsmt & 63 & 58 & 30 & 38 \\
20200911-Pine & 245 & 243 & 239 & 239 \\
20211101-Geogebra & 112 & 110 & 101 & 98 \\
20220314-Uncu & 225 & 224 & 222 & 222 \\
hong & 20 & 12 & 11 & 11 \\
hycomp & 2752 & 2388 & 2426 & 2472 \\
kissing & 45 & 15 & 14 & 14 \\
LassoRanker & 821 & 389 & 571 & 613 \\
meti-tarski & 7006 & 7002 & 6974 & 6960 \\
UltimateAutomizer & 61 & 57 & 52 & 51 \\
zankl & 166 & 87 & 83 & 86 \\\hline
Total & 12134 & 10778 & 10920 & 11005\\
\end{tabular}
}
\vspace{2mm}
\caption{Comparison of solved instances for different branching heuristics.}
\label{tab:effect_prop}
\end{table}

\subsection{Threats to Validity}
\noindent\textbf{Correctness of implementation.} Developing \texttt{clauseSMT} required substantial effort. All comparisons with other solvers were executed in the same environment. Results were carefully verified, and satisfiable instances were validated to ensure correctness.

\noindent\textbf{Randomness.} NLSAT uses internal randomness for semantic decisions and clause/literal reordering. These mechanisms are preserved in our solver, so they do not affect comparisons. Additionally, a fully random literal-decision variant was tested (and found uncompetitive, as discussed). Key metrics such as conflict counts and search stages are reported via scatter plots.

\section{Related Work}
\label{sec:related}
SMT-solving methods can be categorized into complete and incomplete approaches. Incomplete methods are fast due to specialized techniques. Interval constraint propagation (ICP)~\cite{KhanhO12, TungKO17}, as implemented in dReal~\cite{GaoKC13}, is widely used for quickly detecting unsatisfiable instances. Local search has also been extended from SAT to arithmetic theories, including integer~\cite{CaiLZ22, CaiLZ2023}, linear/multilinear real~\cite{multilinear}, and nonlinear real arithmetic~\cite{LiXZ23, WZHNRA}.

Complete methods dominate modern SMT solvers, performing well on both satisfiable and unsatisfiable instances. CDCL(T)~\cite{NieuwenhuisOT06} and NLSAT~\cite{JovanovicM12} rely on CAD for theory reasoning~\cite{Kremer20}, while MCSAT maintains high performance across diverse applications using lighter explanation modules~\cite{MouraJ13}. Recent work has focused on improving NLSAT efficiency, e.g., by optimizing variable projection orders~\cite{LiXZZ23, CADorder2, CADorder3}, designing innovative projection operators~\cite{sample-cell}, generating larger literal-invariant cells~\cite{AbrahamDEK21, levelWise}, and exploring dynamic branching heuristics~\cite{MCSATOrder}.

\section{Conclusion}
\label{sec:conclu}
We presented a clause-level NLSAT algorithm for SMT solving over nonlinear real arithmetic. We categorized conflicts in NLSAT and analyzed the challenges of literal decisions faced by many CDCL-style algorithms. To address these challenges, we introduced a feasible-set-based look-ahead mechanism and clause-level propagation for branching. Experimental results show that our solver is competitive with mainstream solvers, demonstrating the effectiveness of the proposed techniques.

In future work, we aim to develop a clause-level approach for block cases, which are closely related to quantifier elimination. We anticipate that lighter alternatives to CAD may allow connecting literals across clauses, revising previous decisions, and constructing consistent decision paths.

\section{Acknowledgments}
\label{sec:ack}
The author gratefully acknowledges the use of computing resources provided by the Institute of Software, Chinese Academy of Sciences. The author also sincerely thanks the anonymous reviewers for their constructive comments and suggestions, which helped to improve the quality of this paper.

\bibliographystyle{IEEEtran}
\bibliography{clauseSMT}

\end{document}